# Local excitation of surface plasmon polaritons using nitrogen-vacancy centers

Cesar E. Garcia-Ortiz,[1,*] Shailesh Kumar,[2] Sergey I. Bozhevolnyi,[2]

[1]CICESE, Unidad Monterrey, Alianza Centro 501, PIIT Apodaca, Nuevo Leon, 66629, Mexico
[2]Department of Technology and Innovation, University of Southern Denmark, Niels Bohr Allé 1, 5230 Odense M, Denmark
*Corresponding author: cegarcia@cicese.mx

Surface plasmon polaritons (SPPs) are locally excited at silver surfaces using (~100) nm-sized nanodiamonds (NDs) with multiple nitrogen-vacancy (NV) centers (~400). The fluorescence from an externally illuminated (at 532 nm) ND and from nearby NDs, which are not illuminated but produce out-of-plane scattering of SPPs excited by the illuminated ND, exhibit distinctly different wavelength spectra, showing short-wavelength filtering due to the SPP propagation loss. The results indicate that NDs with multiple NV centers can be used as efficient sub-wavelength SPP sources in planar integrated plasmonics for various applications.

Nitrogen-vacancy (NV) centers in diamond are defects in the crystal structure consisting of a nitrogen atom and a lattice vacancy oriented along the [111] direction [1]. NV centers are of interest because they behave as artificial atoms and can be used as single photon sources [2]. Photostability is another important characteristic of NV centers [3]. In addition, NV centers can be created in desired locations in bulk or deterministically moved and positioned in nanodiamonds [4, 5]. In spintronics, these color centers offer the possibility to measure the electronic spin optically at room temperatures [6], thus facilitating the research in quantum information processing [7]. Although many of the interesting characteristics of NV centers are interesting for single NV centers, the study and application of diamond nanoparticles with multiple NV centers is also significant. Nanodiamonds (NDs) with multiple NV centers can be used to measure magnetic fields with higher precision, and with plasmonic devices it can be improved further [8]. Surface plasmon polaritons (SPPs) manifest as evanescent electromagnetic fields at the interface between a metal and a dielectric and can be confined well below the diffraction limit, thus the use of a sub-wavelength, tightly localized optical source to excite SPPs is a natural need [9].

In recent work, NDs with NV centers have been used as single photon sources to couple propagating SPPs in silver nanowires where a reduction of lifetime [5,10], and the wave-particle duality were observed [11]. Additionally, the leakage radiation of SPPs excited with fluorescence from a ND, mounted on a scanning near-field microscopy (SNOM) tip, was measured to show the preservation of coherence through the first and second order correlation experiments [12]. Even though the latter reference deals with more than one NV center (~5), the research is still in the frontier between classical and quantum optics. As commented before, most of the research related to SPPs and NV centers point towards quantum properties and have left the classical approximation unattended.

In this Letter, we address the local excitation of SPPs with sub-wavelength optical sources using a single ND with multiple NV centers. The SPP propagation losses are estimated by comparing the fluorescence spectrum emitted at the source and the spectrum of the light dispersed by a scatterer. The main idea behind such an experiment is to obtain a classical characterization of the NV-SPP coupling, scattering and associated losses, in order to validate their use for more complex plasmonic devices, such as in dielectric-loaded waveguides [13], V-grooves [14], nano-antennas [15], etc.

The NDs used in this work consist of two nanocrystals with diameters around 100 nm and each ND containing around 400 NV centers. One ND was used as the local SPP source (ND1), and a second one (ND2), as a scatterer to out-couple the SPPs. A silica coated silver thin film was fabricated to support the SPPs. The silver layer was 70 nm thick and the protective coating consisted of 15 nm of $SiO_2$ to prevent silver reacting with sulfur compounds as well as its oxidation (Fig. 1a). The diamond nanoparticles were positioned on top of the sample using a technique that requires patterning holes on a PMMA (Polymethyl methacrylate) resist using e-beam lithography and followed by a liftoff process [16]. This patterning was needed in order to ensure that NDs would be positioned individually at a controlled distance, preventing thereby their (uncontrolled) clustering in large lumps. An atomic force microscope (AFM) image showing an ND pattern with a period of 3 µm is presented in Fig. 1b. In Fig. 1c., an SEM image of a single ND is presented. For the experiment presented in this letter, the NDs were separated by a distance $d = 24$ µm. This rather large distance was needed to ensure that one could neglect both direct (far-field) illumination of another ND by freely propagating spherical waves, whose amplitudes decrease much faster than that of SPPs (and are very weak for the propagation direction parallel to the surface), and the contribution associated with the SPP excitation at the pump wavelength (with its subsequent pumping another ND causing the fluorescence of the latter), because the SPP propagation length at the pump wavelength of 532 nm is estimated to be ~ 5 µm. The source ND was pumped using a frequency doubled Nd:YAG laser at a wavelength of 532 nm focused with a 100x objective of an inverted microscope to a 1-µm-wide spot at the sample surface. The fluorescence and scattered light from both NDs was collected with the same objective, and a longpass filter (550 nm) was used to stop the excitation beam from reaching the detectors. An analyzer was used to filter the different field components.

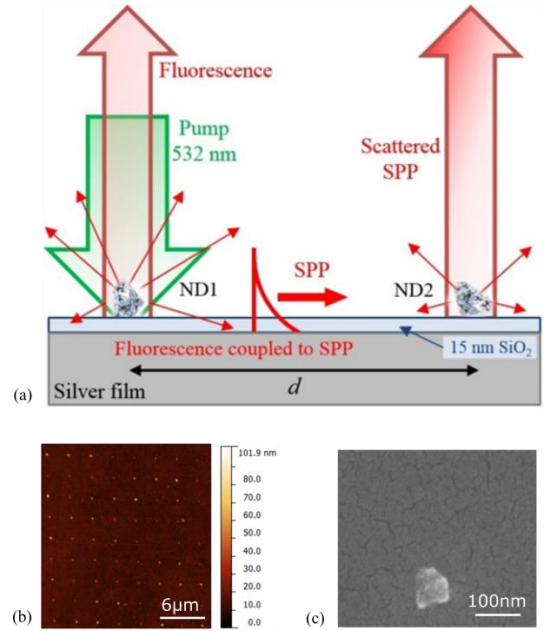

Fig. 1. (a) Schematic diagram of the experimental configuration showing two nanodiamonds separated a distance $d = 24$ μm on a silver film coated with 15 nm of $SiO_2$. ND1 is the local SPP source which excites a SPP propagating in the direction towards ND2, which out-couples the SPP into scattered light. (b) Atomic force microscope (AFM) image of a periodic pattern of NDs. (c) Scanning electron microscope (SEM) image of an ND.

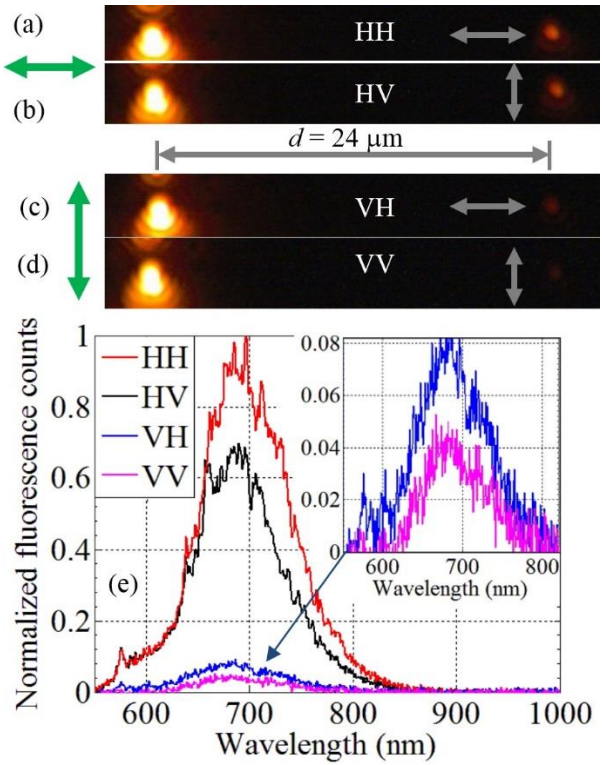
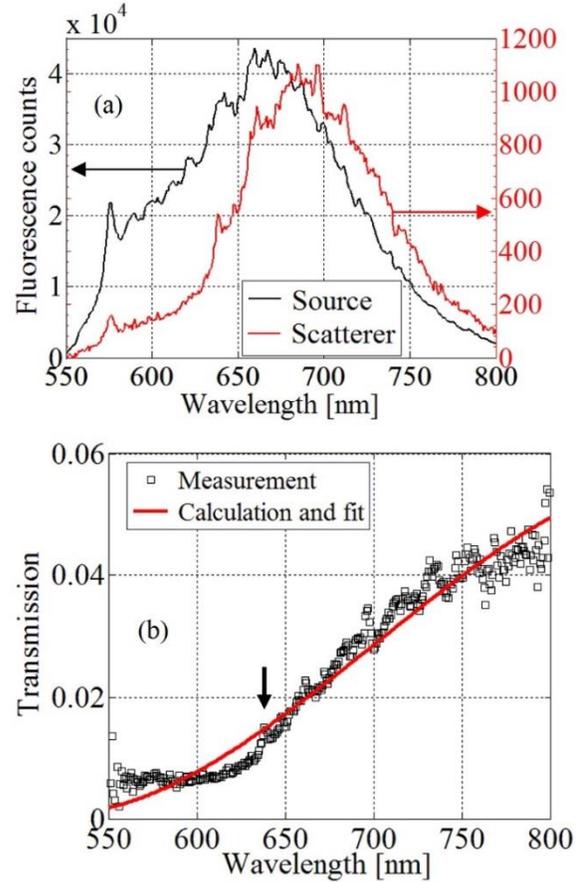

Fig. 2. (a–d) Optical image recorded with the CCD camera showing a bright spot to the left from the fluorescence of local source ND1 and a dimmer spot to the right corresponding to the scattered light from ND2. The arrows in the left represent the excitation polarization and the arrows to the right represent the relative orientation of the analyzer. (e) Spectra of the scattered light emitted by ND2. The inset is a zoom of VH and VV.

Fig. 3. (a) Fluorescence and scattered light spectra from the source ND1 and scatterer ND2. The excitation polarization was in the H-direction. The analyzer was removed for this measurement. (b) Transmitted intensity from ND1 to ND2 obtained from the ratio of the spectra in (a).

A charge-coupled device (CCD) camera in the microscope was used to image the fluorescence spots emitted from the source ND1 and the scatterer ND2 for two polarization directions, perpendicular to each other, filtered with the analyzer [Fig. 2(a–d)]. The polarization direction along the line that connects the particles is labeled as H (horizontal), and the perpendicular direction is labeled as V (vertical) [Fig. 2]. The four possible combinations were analyzed and showed a clear, brighter spot coming from ND2 for the HH configuration [Fig. 2(a)]. Also, as expected, no significant change was observed for any configuration from the source (ND1). SPPs have the strongest field component in the direction of propagation, and it is therefore natural that the HH configuration shows the highest intensity, thus confirming the excitation and propagation of SPPs from ND1 to ND2. For quantitative measurements, we adapted a lens and an iris diaphragm into the experimental setup to filter and select the radiation of a specific spot to analyze with a spectrum analyzer [Fig. 2(e)]. As observed qualitatively, the highest scattered light intensity occurs for the HH configuration, and an intensity decrease of 28% when crossing the analyzer (HV). Importantly, when ND1 was excited with the polarization perpendicular the line between ND1 and ND2 (V-direction), the scattered light intensity arising from ND2 decreased by an order of magnitude.

Such a large difference in the scattered by ND2 fluorescence from ND1 is related to the circumstance that the NV dipoles excited in this perpendicular direction cannot be coupled to SPP waves propagating towards ND2. Therefore, the illumination polarized in the perpendicular direction excites primarily the NV dipoles that do not contribute to the (SPP-mediated) light scattering by ND2, and vice versa. By the same token, the observation of such a strong difference (between HH, HV and VH, VV configurations) indicates that the light scattering by ND2 is indeed mediated via the SPP excitation by ND1. Another important consideration to be taken into account is that the polarization of scattered light, assuming that it is produced by the SPP waves coming from ND1, is expected to be primarily in the (H) direction connecting two NDs – another component in the scattered light can appear only because the ND2 shape is not spherical, which it never is. It is therefore understandable that the highest extinction ratio was found to be HH:VV = 26 (14 dB). Exactly for the same reason, it was found that even when the excitation light polarization was in the V-direction, the highest intensity was detected for the crossed configuration VH. As elucidated above, the reason for this phenomenon is related to the circumstance that the light scattering by ND2 is caused by the SPP waves propagating in the H-direction. In this case, we observe an increase of 230% when crossing the analyzer [inset in Fig. 2(e)].

We use the following model to compare the fluorescence spectrum measured at the illuminated ND, acting as a source of excited SPPs, and that recorded at another ND that scatters the excited SPPs towards a detector. The power spectrum observed at the ND source can be described by $P_{source}(\lambda) = I_g \sigma_{abs} \eta \eta_{rad}(\lambda)$, where $I_g$ is the intensity of excitation laser beam, $\sigma_{abs}$ is the ND absorption cross-section at 532 nm, $\eta$ is the quantum efficiency of NV-centers, and $\eta_{rad}(\lambda)$ is the fraction of photons emitted by NV-centers out of the ND into free-propagating radiation. Similarly, the generated (by the NV centers) SPP power spectrum can be written as: $P_{SPP}(\lambda) = I_g \sigma_{abs} \eta \eta_{SPP}(\lambda)$, where $\eta_{SPP}(\lambda)$ is the fraction of photons coupled to the SPP mode. Then, the SPP intensity at the site of the nearby ND is as follows: $I_{SPP}(\lambda) = P_{SPP}(\lambda)(e^{-d/l_{spp}(\lambda)}/2\pi d a_s)$, where $d$ is distance between the ND source and the ND scatterer, $l_{spp}(\lambda)$ is the SPP propagation length, and $a_s < 1$ is the parameter characterizing the in-plane angular dependence of SPP generation. Finally, the power spectrum generated by the out-of-plane SPP scattering measured at the nearby ND is given by $P_{scat}(\lambda) = I_{SPP}(\lambda)\sigma_{sc}$, where $\sigma_{sc}$ is the SPP-to-light scattering cross-section of the ND scatterer. So, $P_{scat}(\lambda)/P_{source}(\lambda) = f(\lambda) e^{-d/l_{spp}(\lambda)}$, where $f(\lambda) = \eta_{SPP}(\lambda) \sigma_{sc}(\lambda)/\eta_{rad}(\lambda)2\pi d a_s$. The spectrum emitted directly from the source ND1 was measured to compare it with the spectrum of the light scattered by ND2 [Fig. 3(a)]. In this case, the analyzer was removed to capture all field components, and the excitation was in the H-direction. The scattered light from ND2 shows a filtered version compared to the fluorescence spectrum of the local source ND1. SPPs propagate shorter distances for shorter wavelengths, mainly due to ohmic losses in the metal. For this reason, the spectrum of the scattered light is filtered according to the SPP propagation length for each wavelength. This effect can be observed more clearly by examining the transmitted intensity for each wavelength, simply by obtaining the ratio between both spectra [Fig. 3(b)]. This result led to the idea of comparing the measured transmission to a well-known semi-analytical expression of a SPP propagating though a double interface metal-insulator-insulator (MII) system [Fig. 3(b)]. The calculated values do not take into account the coupling and out-coupling efficiency, and therefore the constant efficiency factor $f(\lambda)$ was used as a fitting parameter, yielding $f(\lambda) = 0.12$, which can be considered as a measure of the efficiency of the whole process (of NV coupling to SPP waves propagating from the ND source and being scattered by another remote ND), excluding the propagation losses. The measured and calculated values have a good correspondence ($R^2 = 0.96$), but the calculation fails to fit the values below 637 nm (arrow in Fig. 3(b)]. The dependence of $f(\lambda)$ over wavelength might explain the mismatch. However, the zero-phonon line associated to the negatively charged NV centers (NV-) may also be responsible for part of such effect. Finally, we would like to mention that we obtained qualitatively similar results with other ND pairs located at different distances, whose presentation however would unnecessarily increase the length of the paper and deemed to be superfluous.

In conclusion, NDs with multiple NV centers have shown to be good candidates to be used as highly confined local sources of SPPs, considering that one can obtain much smaller NDs (around tens of nanometers). The high intensity of their fluorescence and photostability allows a systematic use in experiments, for example, in plasmonic waveguides, where reliable, reproducible, and stable SPP sources are needed. The broadband nature of NV fluorescence opens a new possibility to study the plasmonic dispersion without a tunable source and on a single measurement. Moreover, NDs demonstrated to be very efficient scatterers, and thus can be used to probe the SPP intensity at desired points. The effects observed around the zero-phonon line of the NV- (at 637 nm), remains an opportunity area for further research.

The authors gratefully acknowledge the financial support from the European Research Council, grant No. 341054 (PLAQNAP). C.E.G.O. also acknowledges the financial support from CONACYT, and CICESE internal project 692-107.

# References

1. M.W. Doherty, N.B. Manson, P. Delaney, F. Jelezko, J. Wrachtrup, L.C.L. Hollenberg, Phys. Rep. **528** (1), 1 (2013).
2. M. Geiselmann, R. Marty, F.J. García de Abajo, and R. Quidant, Nature Phys. **9**, 785 (2013).
3. S.-J. Yu, M.-W. Kang, H.-C. Chang, K.-M. Chen, and Y.-C. Yu, J. Am. Chem. Soc. **127** (50), 17604 (2005).
4. P Spinicelli , A Dréau , L Rondin , F Silva , J Achard , S Xavier , S Bansropun , T Debuisschert , S Pezzagna , J Meijer , V Jacques1 and J-F Roch, New Journal of Physics 13 (2011) 025014.
5. A. Huck, S. Kumar, A. Shakoor, and U.L. Andersen, Phys. Rev. Lett. **106**, 096801 (2011).
6. D.D. Awschalom, L.C. Bassett, A.S. Dzurak, E.L. Hu, J.R. Petta, Science **339**, 1174 (2013).
7. G.D. Fuchs, G. Burkard, P.V. Klimov and D.D. Awschalom, Nature Phys. **7**, 789 (2011).
8. J. M. Taylor *et. al.* Nature Physics **4**, 810 (2008).
9. M.S. Tame, K.R. McEnery, S.K. Özdemir, J. Lee, S.A. Maier and M.S. Kim, Nature Phys. **9**, 329 (2013).
10. S. Kumar, A. Huck , and U.L. Andersen, Nano Lett. **13** (3), 1221 (2013).
11. R. Kolesov, B. Grotz, G. Balasubramanian, R.J. Stöhr, A.A.L. Nicolet, P.R. Hemmer, F. Jelezko and J. Wrachtrup, Nature Phys. **5**, 470 (2009).
12. O. Mollet, S. Huant, G. Dantelle, T. Gacoin, A. Drezet, Phys. Rev. B **86**, 045401 (2011).
13. Z. Han, C.E. Garcia-Ortiz, I.P. Radko, and S.I. Bozhevolnyi, Opt. Lett. **38** (6), 875 (2013).
14. C.L.C. Smith, A.H. Thilsted, C.E. Garcia-Ortiz, I.P. Radko, R. Marie, C. Jeppesen, C. Vannahme, S.I. Bozhevolnyi, and A. Kristensen, Nano Lett. **14** (3), 1659 (2014).
15. V.A. Zenin, A. Pors, Z. Han, R.L. Eriksen, V.S. Volkov, and S.I. Bozhevolnyi, Opt. Express **22** (9) 10341 (2014).
16. D. Dregely, K. Lindfors, J. Dorfmüller, M. Hentschel, M. Becker, J. Wrachtrup, M. Lippitz, R. Vogelgesang, and H. Giessen, Phys. Status Solidi B **249** (4), 666 (2012).